\documentclass[11pt,a4paper]{article} 
\usepackage{jheppub,hyperref,mdwlist}
\usepackage{jheppub,hyperref,mdwlist}
\usepackage{amsmath}
\usepackage{graphicx}
\usepackage{subfigure}
\usepackage{amssymb}
\usepackage{xcolor}
\usepackage{multirow}
\usepackage{cancel}
\usepackage{color}
\usepackage{mathtools}
\usepackage{listings}
\usepackage{xcolor}
\usepackage{float}
\usepackage{slashed}
\usepackage{bm}
\usepackage{amsmath}
\usepackage{wrapfig}

\newcommand{\be}{\begin{equation}}
\newcommand{\ee}{\end{equation}}
\newcommand{\beq}{\begin{equation}}
\newcommand{\eeq}{\end{equation}}
\newcommand{\bea}{\begin{eqnarray}}
\newcommand{\eea}{\end{eqnarray}}
\definecolor{airforceblue}{rgb}{0.36, 0.54, 0.66}
\definecolor{steelblue}{rgb}{0.27, 0.51, 0.71}
\definecolor{amber}{rgb}{1.0, 0.49, 0.0}

\title{\boldmath Inflation and Dark Matter in the $Z_5$ Model}

\author{XinXin Qi,}
\author{Hao Sun}
\affiliation{Institute of Theoretical Physics, School of Physics, Dalian University of Technology, No.2 Linggong Road, Dalian, Liaoning, 116024, P.R.China }
\emailAdd{qxx@mail.dlut.edu.cn}
\emailAdd{haosun@dlut.edu.cn}

\abstract{
  We discuss the possibility of unifying dark matter physics and inflation in the $Z_5$ model of the two-component dark matter. Inflation driven by the two-component dark matter fields can be divided into two cases, singlet dark matter inflation and mixed dark matter inflation, where both two-component play the role of inflaton in the latter case. 
For dark matter, we focus on the mixed dark matter inflation case. We show a viable parameter space that satisfies  the theoretical and dark matter relic density constraint in the case of successful inflation. It turns out that the dark matter density is dominated by the light component, which is consistent with the feature of the $Z_5$ model of the two-component dark matter.
}

\begin{document}
\maketitle
\flushbottom

\section{Introduction}

The Standard Model~(SM) with $\rm SU(3)_C \otimes SU(2)_L \otimes U(1)_Y$ gauge group has achieved unparalleled success in particle physics, thanks to the discovery of the last missing piece in SM, the Higgs boson~{\cite{ATLAS:2012yve,CMS:2012qbp}}. However, when connecting SM phenomenology with that of modern cosmology, there exist even more outstanding problems unsolved. For example, the assumptions of inflation and dark matter, which are supported by the precise measurement of Cosmic Microwave Background~(CMB) in the Planck~{\cite{Liu:2021diz}} and the WMAP~{\cite{WMAP:2010qai}} experiments.

For the issue of inflation, it works well in answering the horizon problem and homogeneity problem in CMB.
 One general method of constructing an inflation model is to introduce scalar fields coupled or uncoupled with the SM Higgs field. Such an inflationary model can provide initial conditions for the subsequent thermal big bang evolution of the universe. Another important peculiarity is that it can also explain the temperature anisotropy in  CMB and large scale structures in the universe because  the quantum fluctuations of the inflation field are stretched to the cosmic scale during inflation. 
Furthermore, the inflation parameters such as the spectral index, the tensor to scalar ratio and the scalar power spectrum extracted from the experiment~{\cite{Liu:2021diz,WMAP:2010qai}} can restrict the specific inflationary model when the slow roll parameter $\epsilon \approx 1$.
The simplest method is to select the Higgs field in SM as the inflation field~{\cite{Bezrukov:2007ep,Bezrukov:2010jz}} without introducing extra scalar fields, and the only additional parameter comes from the $\xi R |H|^2$ term.
However, there are two problems in such a model:  one is that the quartic self-coupling of the Higgs field $\lambda$ becomes negative at high energy scales~($\gtrsim 10^{10}$ GeV), which destroys the stability of the vacuum~{\cite{Sher:1988mj}}. The other is that at the scale around $M_{Pl}/{\xi} \approx 10^3$\ GeV~{\cite{Burgess:2009ea}}, the unitarity is broken by requiring $\xi \sim 10^4$ due to the scalar power spectrum boundary~{\cite{Lerner_2010}}.
To overcome the unitarity problem at the unitarity scale in the Higgs inflation model, an additional dynamical degree of freedom can be introduced~{\cite{Giudice:2010ka}}. In this case, the extra gauge singlet scalar particle can act as the inflaton and the Higgs field in the SM will play the role of reheating the universe. Another way to remedy the  unitarity-limit problem of the Higgs inflation is that one can introduce $R^2$ term in the Lagrangian (which also introduces another scalar field so-called scalaron in the model), and related discussion can be found in \cite{Gorbunov:2018llf,Bezrukov:2019ylq}.
Generally speaking, people talk more about chaotic inflation or Starobinsky inflation. The former has a power law potential of $m^2\phi^2 + \lambda \phi^4$, and the latter has exponential potential. We will focus on the latter case, and for more details on inflationary cosmology, we refer to the review~{\cite{Linde:2014nna}}.

Dark Matter~(DM) is another outstanding issue, which has been studied extensively for many years. Recently, the Planck experiment~{\cite{Liu:2021diz}} and other astronomical observations put forward a new limit on the abundance of DM, that is $\Omega_{DM}h^2 \approx 0.12$.
The most widely studied dark matter scenario is the so-called weakly interacting massive particles~(WIMP) in which the mass of DM candidate particles ranges from GeV to TeV. Other DM scenarios are also optional, including but not limited to axions, feebly interacting massive particles (FIMP), strongly interacting massive particles (SIMP) and forbidden dark matter~{\cite{Hall:2009bx,JOUR,Baer:2014eja,Hochberg:2014dra}}. As one of the simplest dark matter models, the gauge singlet scalar dark matter model  has been studied for a long time~{\cite{Cline:2013gha,Queiroz:2014yna}}. The more complex DM model  with the scalar dark matter usually carries additional discrete symmetry, which can be regarded as a remnant of the spontaneously broken of some $U(1)$ gauge symmetry, such as $Z_2$ and $Z_N$~{\cite{Batell:2010bp,Belanger:2014bga}}. On the other hand, more studies have shown that there is no reason to assume that the observed dark matter density is contributed just by a single new particle~{\cite{Boehm:2003ha,Ma:2006uv,Cao:2007fy,Hur:2007ur,Lee:2008pc,Zurek:2008qg,Profumo:2009tb,Baer:2011hx,Esch:2014jpa}}. In Ref.{\cite{Belanger:2020hyh}}, a two-component dark matter model based on the $Z_5$ symmetry has been studied. Here the $Z_5$ symmetry serves as the prototype of all $Z_N$ scenarios in which the DM particles are two complex scalars. Since the particles carry different $Z_5$ charges and assume $M_1/2 < M_2 <2M_1$, both the two DM particles are stable and can be selected as WIMP.                                                                                                                                                                                                                                                                                                                                                                                                                                                                                                                                                                                                                                                                                                                                                                                                                                                                                                                                                                                                                                                                                                                                                                                                                                                                                                                                                                                                                                                                                                                                                                                                                                                                                                                                                                                                                                                                                                                                                                                                                                                                                                                                                                                                                                                                                                                                                                                                                                                                                                                                                                                                                                                                                                                                                                                                                                                                                                                                                                                                                                                                                                                                                                                                                                                                                                                                                                                                                                                                                                                                                                                                                                                                                                                                                                                                                                                                                                                                                                                                                                                                                                                                                                                                                                                                                                                                                                                                                                                                                                                                                                                                                                                                                                                                                                                                                                                                                                                                                                                                                                                                                                                                                                                                                                                                                                                                                                                                                                                                                                                                                                                                                                                                                                                                                                                                                                                                                                                                                                                                                                                                                                                                                                                                                                                                                                                                                                                                                                                                                                                                                                                                                                                                                                                                                                                                                                                                                                                                                                                                                                                                                                                                                                                                                                                                                                                                                                                                                                                                                                                                                                                                                                                                                                                                                                                                                                                                                                                                                                                                                                                                                                                                                                                                                                                                                                                                                                                                                                                                                                                                                                                                                                                                                                                                                                                                                                                                                                                                                                                                                                                                                                                                                                                                                                                                                                                                                                                                                                                                                                                                                                                                                                                                                                                                                                                                                                                                                                                                                                                                                                                                                                                                                                                                                                                                                                                                                                                                                                                                                                                                                                                                                                                                                                                                                                                                                                                                                                                                                                                                                                                                                                                                                                                                                              More works have recently attempted to combine inflation with dark matter in one framework. In Refs.~{\cite{Kahlhoefer_2015,Lerner:2009xg,Aravind:2015xst}}, a gauge singlet scalar acts as the inflaton and then acts as a dark matter candidate in the freeze-out scenario.
Ref.{\cite{Okada:2010jd}}
studied a scalar WIMP dark matter candidate with non-minimal coupling to gravity, where the scalar dark matter acts as the inflaton.
Ref.{\cite{Gong:2012ri}} discussed  inflation with the two Higgs doublet model at low energy. In ref.{\cite{Kim:2014kok}}, realizations of the SM Higgs inflation in the presence of Higgs-portal interactions are studied, and the author discussed connections between singlet fermion dark matter  and inflation.
In ref.{\cite{Choubey:2017hsq}}, the reheating after inflation was discussed in detail in the $Z_2$ dark matter model
Finally, a scenario linking inflation and dark matter in the model with gauged $Z_3$ symmetry was discussed in ref.{\cite{Choi:2020ara}}. 

In our present paper, we connect the slow-roll inflation with the non-minimal gravity couplings to the two-component dark matter model with $Z_5$ symmetry. Due to non-minimal gravity couplings, more than one scalar field beyond the SM exists, which will lead to a pure field or multi-field inflation. They can then play the role of DM in the universe during cooling down. According to \cite{Almeida:2018oid}, an electrically neutral inflaton scalar field stable on the cosmological scale can both drive the cosmic inflation and  constitute a dark matter (DM) component, this can happen either because the reheating stage was not complete after inflation and left behind a remnant of the scalar condensate, or due to the fact that the inflaton excitations were eventually created by the SM production following the decay of the inflaton zero mode. In this work, we consider that decays of the homogeneous inflaton condensate are always complete after inflation. Since reheating is a non-perturbative process where a time-dependent condensate transfers its energy density into other fields \cite{Kofman:1994rk,Mukaida:2013xxa}, the inflaton particles can still be stable. We consider the
couplings between the inflaton (DM) and the SM sector are  not too small, the inflaton particles (DM) can therefore
enter in thermal equilibrium with the other particles produced in reheating. Then, the stable inflaton particles will undergo thermal freeze-out and constitute the usual thermal relic. We also discuss the theory constraints and inflation dynamics  of the $Z_5$ dark matter model in this work.

Our paper is organized as follows. In Sec.\ref{sec2:model}, we introduce the model framework. In Sec.\ref{sec:constraints}, we study the inflation dynamics in the cases of singlet dark matter inflation and mixed dark matter inflation. In Sec.\ref{z5c}, we discuss the constraints on the model, including unitarity, perturbativity and vacuum stability. We 
 discuss the DM phenomenology in the model with some fixed parameters in Sec.\ref{dm}. Finally, we give our conclusion in Sec.\ref{sec:conclusions}.

\section{The $Z_5$ model of two-component dark matter for inflation}
\label{sec2:model}

The general two-component dark matter field action under $Z_5$ symmetry with the Jordan frame is given by
\begin{eqnarray}\nonumber
\label{acJD}
\frac{{\cal L}_J}{\sqrt{-g_J}} &=& -\frac{1}{2}M^2_{Pl}R -\Big(\xi_0|H|^2+\xi_1|\Phi_1|^2+\xi_2|\Phi_2|^2\Big)R
-\left|{D_\mu}H\right|^2-\left|D_\mu\Phi_1\right|^2-\left|D_\mu\Phi_2\right|^2 \\
&-&{{\cal V}_J}\left(H,\Phi_1,\Phi_2\right)\ ,
\end{eqnarray}
where the general renormalizable potential is:
\begin{align}
\label{eq:Z5lag}
{{\cal V}_J}\left(H,\Phi_1,\Phi_2\right)&=\,\,-\mu^2_H|H|^2+\lambda_H|H|^4+\mu_{1}^2|\Phi_1|^2+\lambda_{41}|\Phi_1|^4+\lambda_{S1}|H|^2|\Phi_1|^2\nonumber\\
 & \,+\mu_{2}^2|\Phi_2|^2+\lambda_{42}|\Phi_2|{^4}+\lambda_{S2}|H|^2|\Phi_2|^2+\lambda_{412}|\Phi_1|^2|\Phi_2|^2\nonumber\\
 & \,+\frac{1}{2}\left[\mu_{S1}\Phi^2_1\Phi_2^{*} + \mu_{S2}\Phi_2^2\Phi_1 +\lambda_{31}\Phi _1^3 \Phi _2+\lambda_{32}\Phi _1 \Phi _2^{*3} + \text{h.c.}\right].
\end{align}
The SM Higgs doublet and the components of the two dark matter complex singlets are:
\begin{equation}
H=\frac{1}{\sqrt{2}}\left(\begin{array}{l} 0 \\ h \end{array} \right) \, , \quad
\Phi_1=\frac{1}{\sqrt{2}}\phi_1e^{i\theta_1}\, ,\quad
\Phi_2=\frac{1}{\sqrt{2}}\phi_2e^{i\theta_2}\, .\quad
\end{equation}
In Eq.(\ref{acJD}), $M_{Pl}$ represents the reduced Planck mass ($M_{Pl} \equiv 1/\sqrt{8\pi G_N} = 2.44\times 10^{18} $~GeV), which will be set by 1 later but will be restored if necessary. $R$ is the Ricci scalar. For the  non-minimal dimensionless couplings, we label them as $\xi_i$ ($i=0,1,2$) and require their values to be positive to ensure potential stability under large field values. $H$ is the SM Higgs doublet field and will obtain the vacuum expectation value~{(vev)} {$v$} after spontaneously symmetry breaking. We consider the vevs of the two-component dark matter fields to be zero under the $Z_5$ symmetry and the fields are labeled as $\Phi_1$, $\Phi_2$. $D_{\mu}$ stands for the covariant derivative and can be treated as the normal derivative at the inflation epoch: $D_{\mu}\to\partial_{\mu}$. The minus sign in the kinetic terms is needed to obtain the metric convention of $(-,+,+,+)$. $\mathcal{V}_J$ is the  general $Z_5$-invariant scalar potential~\cite{Belanger:2020hyh}. We have the  transform of $SM \to SM$, $\Phi_1 \to e^{i2\pi/5}\Phi_1$ and $\Phi_2 \to e^{i4\pi/5} \Phi_2$ under the $Z_5$ symmetry. The new trilinear and quartic couplings denoted $\mu_{Si}$ and $\lambda_{3i}~({i= 1,2})$ respectively, can be found in the model. As we can see, with the simultaneous disappearance of the $\mu_{Si}$ and $\lambda_{3i}$ terms, the scalar potential with $Z_5$ symmetry is indistinguishable from that  $Z_2\times Z^{'}_2$ symmetry. In addition, by choosing the phase of $\Phi_{1,2}$, the parameter $\lambda_{3i}$ can be real or complex. For simplicity, we assume that all the parameters in Eq.(\ref{eq:Z5lag}) are real.

We assume that $h= 0$ initially and neglect all terms containing $h$ according to \cite{Almeida:2018oid,Lebedev:2021zdh}. Then, the inflation dynamics can be driven by $\Phi_1$ or $\Phi_2$, or both. 

To eliminate the terms containing $R$ and $\Phi_i$ in Eq.(\ref{acJD}), we perform a conformal transformation from the Jordan frame to the Einstein frame. Note that the parameters associated with conformal transformation will be equal to 1, making the Jordan and Einstein frame equivalent after the end of inflation. 
The conformal transformation of the metric $g_J$ is given by~\cite{Kaiser:2010ps}:
\begin{eqnarray}
g^J_{\mu\nu}=g^E_{\mu\nu}/\Omega^2  \end{eqnarray}
with
\begin{eqnarray}
\Omega^2\equiv 1+\xi_1\phi_1^2+\xi_2{\phi}_2^2.
\end{eqnarray}
Then, we get the Einstein frame action:
\begin{equation}
\frac{{\cal L}_E}{\sqrt{-g_E}}= -\frac{R}{2}-\frac{1}{2}G_{ij}\partial_\mu \phi^i\partial^\mu\phi^j-{{\cal V}_E}(\phi_1,\phi_2),
\end{equation}
where:
\begin{align}
G_{ij}&=\frac{1}{\Omega^2}\delta_{ij}+\frac{3}{2}\frac{1}{\Omega^4}\frac{\partial\,\Omega^2}{\partial\,\phi_i}\frac{\partial\,\Omega^2}{\partial\,\phi_j},\\ 
{\cal V}_E&=\frac{{\cal V}_J}{\Omega^4}.
\end{align}
Here, $G_{i,j}$ is a $2 \times 2$ matrix, ${\cal V}_E$ denotes the scalar potential in Einstein frame, the phase $\theta_1$ and $\theta_2$ can set to be zero by minimizing the scalar potential at $\phi_1 \neq 0$, $\phi_2 \neq 0$. In the large-field limit $\xi_1\phi^2_1 + \xi_2\phi^2_2 \gg 1$, we can drop the scalar masses and the cubic terms, then $G_{i,j}$ and ${\cal V}_E$  are unfolded as:
\begin{align}
G_{i,j} = & \frac{1}{1+\xi_1\phi^2_1+\xi_2 \phi^2_2}
\begin{pmatrix}
1+\dfrac{6\xi^2_1 \phi^2_1}{1+\xi_1\phi^2_1+\xi_2 \phi^2_2} & \dfrac{6\xi_1\xi_2\phi_1\phi_2}{1+\xi_1\phi^2_1+\xi_2 \phi^2_2} 
\\
\dfrac{6\xi_1\xi_2\phi_1\phi_2}{1+\xi_1\phi^2_1+\xi_2 \phi^2_2} & 1+\dfrac{6\xi^2_2 \phi^2_2}{1+\xi_1\phi^2_1+\xi_2 \phi^2_2}  \\
\end{pmatrix} \, ,
\\
{\cal V}_E = & \frac{\lambda_{41}\phi_1^4 + \lambda_{42}\phi_2^4 + \lambda_{412}\phi_1^2\phi_2^2 - \lambda_{31}\phi_1^3\phi_2 - \lambda_{32}\phi_1\phi_2^3}{4\left(1+\xi_1\phi^2_1+\xi_2 \phi^2_2 \right)^2} \, .
\label{epot2}
\end{align}
With the non-zero mixed kinetic term $G_{1,2}$, the diagonal kinetic form can be obtained by redefining
\begin{align}
\varphi= & \sqrt{\frac{3}{2}}\log(1+\xi_1 \phi^2_1+\xi_2 \phi^2_2) \, ,  \label{newfield1}
\\
r = & \frac{\phi_2}{\phi_1} \, . \label{newfield2}
\end{align}
The kinetic term and scalar potential in the Einstein frame can be rewritten as:
\begin{align}
\mathcal{L}_{{\rm kin}} &=
\frac{1}{2}
\frac{(\xi_{1}^{2}+\xi_2^{2}r^{2})}
{(\xi_{1} + \xi_2r^{2})^{3}}\left(
1 - e^{-\sqrt{\frac{2}{3}}\varphi}
\right)
({\partial}r)^2
 +\frac{(\xi_{1}-\xi_2)r}
{\sqrt{6}(\xi_{1}+\xi_2r^{2})^{2}}
({\partial}\varphi)({\partial}r)
\nonumber\\
&\quad+\frac{1}{2}\left(
\frac{e^{\sqrt{\frac{2}{3}}{\varphi}}(1+6\xi_{1}+(1+6\xi_2)r^{2})-6(\xi_{1}+\xi_2r^{2})}
{6(\xi_{1}+\xi_2r^{2})(e^{\sqrt{\frac{2}{3}}{\varphi}}-1)}\right)
({\partial}\varphi)^2\,,\\
{\cal V}(r,\varphi) &={\cal V}(r)\left(1 - e^{-\sqrt{\frac{2}{3}}{\varphi}}\right)^{2}\,,\label{StaP}
\end{align}
with
\begin{equation}\label{Vr}
{\cal V}(r)=  \frac{\lambda_{41} - \lambda_{31}r+ \lambda_{412}r^2- \lambda_{32}r^3+\lambda_{42}r^4}{4\left(\xi_2 r^2+\xi_1\right)^2}.
\end{equation}
The kinetic terms can be reduced further by introducing a canonical scalar field\cite{Choi:2020ara}:
\begin{equation}\label{drcdr}
(\frac{dr_c}{dr})^2=\frac{(\xi^2_1+\xi^2_2r^2)}{(\xi_1+\xi_2r^2)^3},
\end{equation}
Then,
\begin{align}
\mathcal{L}_{{\rm kin}} &\simeq 
\frac{1}{2}({\partial}\varphi)^{2}
+\frac{1}{2}\frac{(\xi_{1}^{2}+\xi_{2}^{2}r^{2})}{(\xi_{1}+\xi_{2}r^{2})^{3}}
({\partial}r)^{2}
+\frac{(\xi_{1}-\xi_{2})r}{\sqrt{6}(\xi_{1}+\xi_{2}r^{2})^{2}}
({\partial}\varphi)({\partial}r)
\nonumber \\
& \simeq
\frac{1}{2}({\partial}\varphi)^{2}
+\frac{1}{2}({\partial}r_c)^{2} 
+({\partial}\varphi)({\partial}r_c)
\frac{(\xi_{1}-\xi_{2})r}
{\sqrt{6}\sqrt{\xi_{1}^{2}+\xi_{2}^{2}r^{2}}
\sqrt{\xi_{1}+\xi_{2}r^{2}}}
\,.
\end{align}
The kinetic mixing term $({\partial}\varphi)({\partial}r)$ can be canceled by setting $\xi_1 \approx \xi_2$
or suppressed by $1/\sqrt{\xi_1 +\xi_2r^2}$ with large non-minimal coupling. Then, the effective action on inflation can be obtained in the form of  vanishing non-minimal couplings and kinetic mixing terms :
\begin{align}
\frac{{\cal L}_{E}}{\sqrt{-g_E}} = 
-\frac{1}{2}{\mathcal{R}}
-\frac{1}{2}({\partial}\varphi)^{2}
-\frac{1}{2}({\partial}r_c)^{2}
-{\cal V}(r_c,\varphi)\,.
\end{align}
Note that the potential in Eq.(\ref{Vr}) has a symmetric exchange between $\Phi_1$ and $\Phi_2$, resulting in $r \rightarrow 1/r$  of nonzero values $\lambda_{31}$ and $\lambda_{32}$, and also obtaining unequal values of non-minimal couplings between $\xi_1$ and $\xi_2$. In the inflation epoch, this feature can provide two distinguishable regions of $\xi_2/\xi_1$ value, {i.e.}, $\xi_2 \approx \xi_1$ and $\xi_2 < \xi_1$.

Now, we discuss the stability of  single-field inflation affiliated with the potential minima(\ref{Vr}) at a fixed point $r_m$. The condition for ${\cal V}(r_m)$ acting as an extreme minimum, requiring that:
\begin{align}
{\cal V}(r)\bigg\vert_{r=r_m} \geq 0
\,,\quad
\frac{\partial {\cal V}(r)}{\partial r_c}\bigg\vert_{r=r_m}
=0
\,,\quad
\frac{\partial^2 {\cal V}(r)}{\partial r_c^2}\bigg\vert_{r=r_m} \geq 0
\,.
\end{align}
Combining Eq.(\ref{Vr}) and Eq.(\ref{drcdr}), the first and second derivatives of the inflation potential are given respectively by:
\begin{align}\label{dvr}
\frac{\partial {{\cal V}(r)}}{\partial r_c} \approx &\frac{M^3_{Pl}}{4(\xi_1+\xi_2r^2)^{3/2}\sqrt{\xi_1^2+\xi_2^2r^2}}[\xi_2 \lambda_{32} r^4+r^3 (4 \xi_1 \lambda_{42}-2 \xi_2 \lambda_{412})+r^2 (3 \xi_2 \lambda_{31}-3 \xi_1 \lambda_{32})\nonumber\\
&+r (2 \xi_1 \lambda_{412}-4 \xi_2 \lambda_{41})-\xi_1 \lambda_{31}].
\end{align}
and
\begin{align}\label{dvr2}
\frac{\partial^2 {{\cal V}(r)}}{\partial r^2_c} \approx &
\frac{M^2_{Pl}}{4(\xi_1+\xi_2r^2)(\xi_1^2+\xi_2^2r^2)^2}
[r^6 \left(2 \xi_2^4 \lambda_{412}-4 \xi_1 \xi_2^3 \lambda_{42}\right)+r^5 \left(\xi_1^2 \xi_2^2 \lambda_{32}+9 \xi_1 \xi_2^3 \lambda_{32}-6 \xi_2^4 \lambda_{31}\right)\nonumber\\
&+r^4 \left(8 \xi_1^2 \xi_2^2 \lambda_{42}-10 \xi_1 \xi_2^3 \lambda_{412}+12 \xi_2^4 \lambda_{32}\right)+r^2 \left(12 \xi_1^4 \lambda_{42}-10 \xi_1^3 \xi_2 \lambda_{412}+8 \xi_1^2 \xi_2^2 \lambda_{32}\right)\nonumber\\
&+r \left(-6 \xi_1^4 \lambda_{32}+9 \xi_1^3 \xi_2 \lambda_{31}+\xi_1^2 \xi_2^2 \lambda_{31}\right)+\left(2\xi_1^4 \lambda_{412}-4 \xi_1^3 \xi_2\lambda_{41}\right)\nonumber\\
&+r^3 \left(7 \xi_1^3 \xi_2 \lambda_{32}-3 \xi_1^2 \xi_2^2 \lambda_{31}-3 \xi_1^2 \xi_2^2 \lambda_{32}+7 \xi_1 \xi_2^3 \lambda_{31}\right)].
\end{align}

\section{Inflation driven by the two-component dark matter fields}
\label{sec:constraints}

We focus on inflationary dynamics driven by one dark matter field inflation or two dark matter field. For the inflation driven by one dark matter field $\Phi_1$ or $\Phi_2$, the Eq.(\ref{dvr}) indicates it can not work in $r \rightarrow 0$ and $r \rightarrow \infty$ scenarios only if $\xi_1\lambda_{31} =0$ and $\xi_2\lambda_{32} =0$ respectively. The mixed inflation makes difference under a finite $r$ scenarios where $r$ makes the potential minimum. The finite $r$ can be obtained by solving the quartic equation in Eq.(\ref{dvr}). We will discuss each $r$ scenario in detail later.

\subsection{Singlet dark matter inflation}

The $r \rightarrow 0$ and $r \rightarrow \infty$ scenarios require $\xi_1 \lambda_{31} = 0$ and $\xi_2\lambda_{32} =0$ respectively according to Eq.(\ref{dvr}). For reducing parameters space, we only consider two cases: A $(\xi_1 \neq 0,\lambda_{31}=0)$ with $r \rightarrow 0$ and B $(\xi_2 \neq 0,\lambda_{32}=0)$ with $r \rightarrow \infty$. The minimal potential condition of $r \rightarrow 0$ or $r \rightarrow \infty$ scenario can be found in the second derivative potential Eq.(\ref{dvr2}) and demands that:
\begin{equation}
\xi_{1}^4\lambda_{412} -2\xi_{1}^3\xi_{2}\lambda_{41} > 0
\end{equation}
or
\begin{equation}
\xi_{2}^4\lambda_{412} -2\xi_{2}^3\xi_{1}\lambda_{42} > 0
\end{equation}

\subsection{Mixed dark matter inflation}
The mixed dark matter inflation make difference when a finite $r$ makes the inflation potential minimum and can be obtained by solving the quartic equation in Eq.(\ref{dvr}) with the nonzero value of $\xi_2$ and $\lambda_{32}$. Generally speaking, the discriminant and solutions expressions of the quartic equation are too long and we do not present them here. For simplicity, we  will add on the conditions as follows:
\begin{align}
4 \xi_1 \lambda_{42}-2 \xi_2 \lambda_{412}&=0\,\label{casec1} ,
\\ 
3 \xi_2 \lambda_{31} - 3\xi_1\lambda_{32}&=0\,\label{casec2},\\
2 \xi_1 \lambda_{412}-4 \xi_2 \lambda_{41}&=0\,\label{casec3} .
\end{align}
Now the solutions of the quartic equation can be expressed as $r_0=(\xi_1\lambda_{31}/{\xi_2\lambda_{32}})^{1/4} $ and we label this scenario as case C.

We unify three cases in one frame nominally, at the inflation direction, the Jordan-frame action can be written as follows:
\begin{align}
S &= \int d^{4}x \, \sqrt{-g} \, \bigg\{
-\frac{M_{{Pl}}^{2}}{2}\left[
1+\left(
\xi_1 + \xi_2 r_m^{2}
\right)\frac{\phi_1^{2}}{M_{{Pl}}^{2}}
\right]\mathcal{R}
\nonumber \\
&\quad
-\frac{1+r_m^{2}}{2}g^{\mu\nu}
\partial_{\mu}\phi_1\partial_{\nu}\phi_1
-\frac{1}{4}(\lambda_{41}-\lambda_{31}r_m - \lambda_{32}r_m^3 +\lambda_{42}r_m^4)\phi_1^4
\bigg\}
\,.
\end{align}
After redefining $\Phi \equiv \sqrt{1+r_m^2}\, \phi_1$, the above action becomes

\begin{align}
S = \int d^{4}x \, \sqrt{-g} \, \left\{
-\frac{M_{{Pl}}^{2}}{2}\left(
1+\xi_\Phi\frac{\Phi^{2}}{M_{{Pl}}^{2}}
\right)\mathcal{R}
-\frac{1}{2}g^{\mu\nu}
\partial_{\mu}\Phi\partial_{\nu}\Phi
-\frac{1}{4}\lambda_\Phi\Phi^4
\right\}
\,,
\end{align}
where
\begin{align}
\xi_\Phi \equiv
\frac{(\xi_1 + \xi_2 r_m^{2})}
{1+r_m^2}
\,,\qquad
\lambda_\Phi \equiv
\frac{(\lambda_{41}-\lambda_{31}r_m - \lambda_{32}r_m^3 +\lambda_{42}r_m^4)}
{(1+r_m^2)^2}
\,.
\end{align}
We end the part with a table[\ref{Tab1}].

\begin{table}[htpb]
    \centering
    \begin{tabular}{c|ccc}
    \hline
    \hline
      & $\xi_\Phi$ & $\lambda_\Phi$ & ${\cal V}(r)$\\
    \hline
        Case A&$\xi_1$&$\lambda_{41}$&$\frac{M^4_{Pl}\lambda_{41}}{4\xi^2_1}$\\
        Case B&$\xi_2$&$\lambda_{42}$&$\frac{M^4_{Pl}\lambda_{42}}{4\xi^2_2}$\\
        Case C&$\frac{(\xi_1 + \xi_2 r_0^{2})}
{1+r_0^2}$&$\frac{(\lambda_{41}-\lambda_{31}r_0 +\lambda_{412}r_0^2-\lambda_{32}r_0^3 +\lambda_{42}r_0^4)}
{(1+r_0^2)^2}$&$\frac{M^4_{Pl}(\lambda_{41} - \lambda_{31}r_0+ \lambda_{412}r_0^2- \lambda_{32}r_0^3+\lambda_{42}r_0^4)}{4\left(\xi_2 r_0^2+\xi_1\right)^2}$\\
     \hline
     \hline
    \end{tabular}
    \caption{The effective parameters $\xi_{\Phi}$, $\lambda_{\Phi}$  and vacuum energy $\mathcal{V}(r)$ in different cases. We do not expand terms with Eq.(\ref{casec1})-Eq.(\ref{casec2}) and $r_0$ just for concision}
    \label{Tab1}
\end{table}


\subsection{Inflationary observables}

The potential given in Eq.(\ref{StaP})  belongs to the so-called Starobinsky potentials{\cite{Starobinsky:1979ty,Calmet:2016fsr}}, as we can see  ${\cal V}(r, \varphi)/{\cal V}(r)$ exhibits flat nature at high field values ensuring slow roll when $r$ is determined by cases. The slow-roll parameters are given by \cite{Baumann:2009ds}:
\begin{align}
\epsilon = \frac{M_{{Pl}}^{2}}{2}\left(
\frac{{\cal V}(r,\varphi)^{\prime}}{{\cal V}(r,\varphi)}
\right)^{2}
\,,\quad
\eta = M_{{Pl}}^{2}\frac{{\cal V}(r,\varphi)^{\prime\prime}}{{\cal V}(r,\varphi)}\quad.
\end{align}
where the prime denotes the derivative with respect to the canonically normalized field  $\varphi$, which is defined by~\cite{Bezrukov:2007ep}:
\begin{align}
\frac{d\varphi}{d \Phi}=\frac{\sqrt{1+(1+6\xi_{\Phi})\xi_{\Phi}\Phi^2/{M^2_{Pl}}}}{1+\xi_{\Phi}\Phi^2/{M^2_{Pl}}} .   
\end{align}
With the help of slow-roll parameters, we can obtain the spectral index $n_s$, the tensor-to-scalar ratio $r$ and the magnitude of the scalar power spectrum $A_s$ at the horizon exit:
\begin{align}
n_s \approx
1-6\epsilon + 2\eta,
r \approx
16\epsilon,
A_s = \frac{{\cal V}(r,\varphi)}{24\pi^4 M^2_{Pl} \epsilon}.
\end{align}

The inflation ends at $\epsilon \approx 1$, and the size of the universe had expanded by $e$ times. The number of $e$-folds $N$ can be obtained as follows:
\begin{align}
N=\frac{1}{M^2_{Pl}}\int_{\varphi_e}^{\varphi_i}d\varphi\frac{{\cal V}(r,\varphi)}{{\cal V}^{\prime}(r,\varphi)}.    
\end{align}
where $\varphi_e$ ($\varphi_i$) is the canonically normalized field value at the end of inflation (at the horizon exit) \cite{Choi:2020ara}. The $\varphi_e$ can be obtained by setting slow roll inflation to the end when $\epsilon \approx 1$ and 
we can compute $\varphi_i$ if $N$ is known. For example, we take $N=60$ at the horizon exit for the typical reheating scenarios , $\varphi_e$ can be estimated as $M_{Pl}(4/3)^{1/4}/\sqrt{\xi_{\Phi}}$.
Furthermore, we express $\varphi_i$ in terms of $N$ as 
\begin{align}
    \varphi_i(N)\approx \sqrt{\frac{4N}{3\xi_{\Phi}}} M_{Pl}
\end{align}
Then, in the large-field limit, $\epsilon$, $\eta$,  $n_s$, $r$ and $A_s$ can be expressed in terms of the number of $e$-foldings,
\begin{align}
\epsilon &\approx \frac{3}{4N^2}
\,,\quad
\eta \approx -\frac{1}{N} \quad.\\
n_s &\approx 1-\frac{2}{N}-\frac{9}{2N^2}
\,,\quad
r \approx \frac{12}{N^2}
\,,\quad
A_s = (\frac{N^2}{72{\pi}^2})\frac{\lambda_{\Phi}}{\xi_{\Phi}^2}.
\end{align}
Recent experiments have revealed the following limitations:

\begin{itemize}
\item CMB normalization  \vspace{0.2cm} \\
The scalar power spectrum at the Planck pivot scale \cite{Planck:2018jri} is given by
\begin{align}
\ln(10^{10} A_s) = 3.044 \pm 0.014
\quad
\text{(68\% C.L. Planck TT,TE,EE+lowE+lensing)}
\,.
\end{align}
Then, requiring $A_s \approx 2.1\times 10^{-9}$ at $N=60$, we obtain the relation between the effective quartic and non-minimal couplings as follows:
\begin{align}
\frac{\lambda_\Phi}{\xi_\Phi^2}
\approx 4.15 \times 10^{-10}
\,.
\end{align}
For instance, for $\lambda_\Phi \simeq 0.1$ ($10^{-7}$), the non-minimal coupling should take $\xi_\Phi \simeq 15500$ ($15.5$).

\item Spectral index and tensor-to-scalar ratio \vspace{0.2cm} \\
The latest Planck data \cite{Planck:2018jri} read
\begin{align}
n_s &= 0.9659 \pm 0.0041 \,&\text{(68\% C.L. Planck TT,TE,EE+lowEB+lensing)}
\,,\\
r &< 0.11 \,&\text{(95\% C.L. Planck TT,TE,EE+lowEB+lensing)}
\,.
\end{align}

\end{itemize}

\begin{figure}[tbp]
\centering
\includegraphics[width=0.45\linewidth]{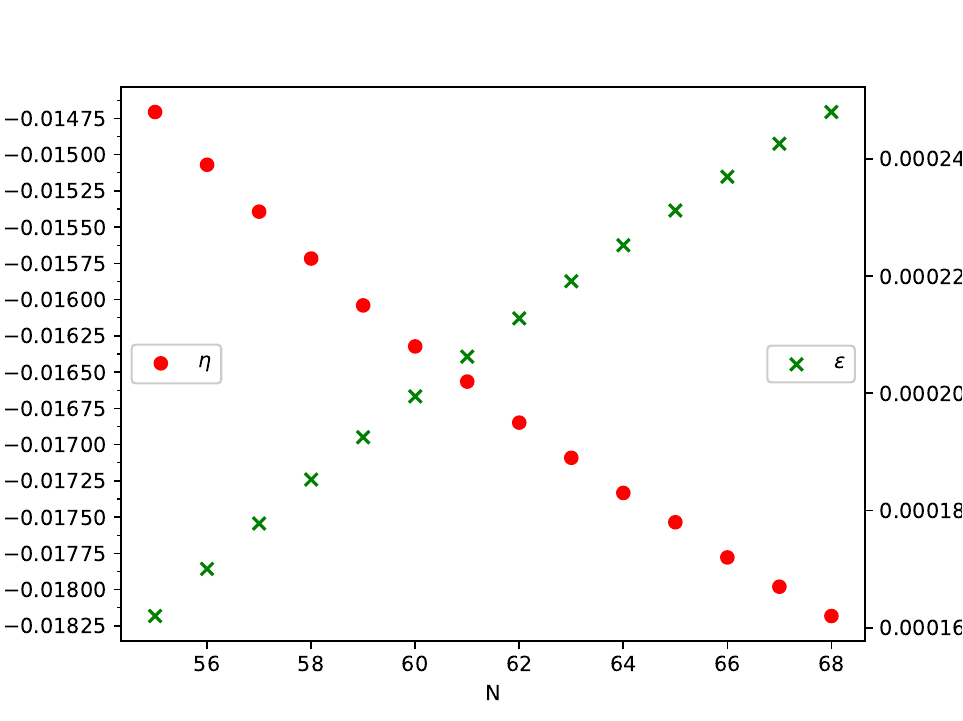}\,\,
\includegraphics[width=0.45\linewidth]{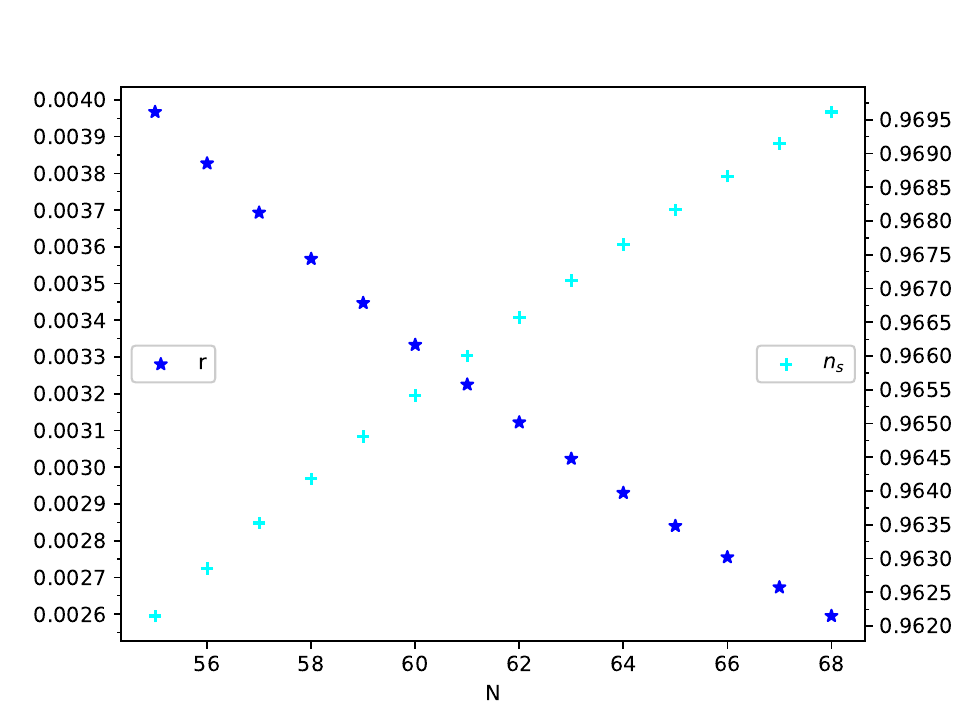}
\caption{The inflationary observables $\eta$, $\epsilon$, $n_s$, $r$ changed by $N$ according to the lastest Planck data\cite{Planck:2018jri}.}
\label{infobs}
\end{figure}

We exhibit inflationary observables changed by the number of $e$-foldings $N$ in Fig.(\ref{infobs}) and talk about $A_s$ in section \ref{infAs}.

\section{$Z_5$ models and low energy constraints}\label{z5c}

\subsection{Unitarity}
The unitarity conditions come from the tree-level scalar-scalar scattering matrix which is dominated by the quartic contact interaction. The s-wave scattering amplitudes should lie under the perturbative unitarity limit, given the requirement the eigenvalues of the S-matrix $M$ must be less than the unitarity bound given by
\begin{align}
|ReM|<\frac{1}{2}.
\end{align}
We extend those scattering matrix by charge and $Z_5$ change conservation states and write down dominated conditions as:
\begin{align}
&\frac{1}{2}\left|(2\lambda_{41}+\lambda_{412})\pm\sqrt{18\lambda^2_{31}+4\lambda^2_{41}-4\lambda_{41}\lambda_{412}+\lambda^2_{412}}\right| < 8\pi,\\
&\frac{1}{6}\left|a_2\pm2\sqrt{a_2^2-3a_{1}}\right| < 8\pi.\\
\end{align}
with
\begin{align}
& a_1=4(16\lambda_{41}\lambda_{42}-\lambda^2_{412}+24\lambda_{41}\lambda_{H}+24\lambda_{42}\lambda_{H}-2\lambda_{S1}^2-2\lambda_{S2}^2),\\
& a _2=4(2\lambda_{41}+2\lambda_{42}+3\lambda_{H}).
\end{align}
More details can be found in Appendix {\ref{uncon}}.

\subsection{Perturbativity}
We need perturbativity on the new dark matter quartic couplings,
\begin{equation}
|3\lambda_{31}|<4\pi,|4\lambda_{41}|<4\pi|,|\lambda_{412}|<4\pi,|3\lambda_{32}|<4\pi,|4\lambda_{42}|<4\pi.
\end{equation}
\subsection{Vacuum stability}
The vacuum stability requires the potential under the restriction,i.e,$\mathcal{V} > 0$ for large field values. We only consider the quartic couplings for their overwhelming contribution to $\mathcal{V}$. In this section, we focus on the vacuum stability induced by dark matter scalars and ignore the mixing quartic couplings with the SM Higgs, this scenario can  work when the mixing quartic couplings with the SM Higgs take positive or small values
as compared to couplings in the dark matter fields.
The potential with vacuum stability can be written as:
\begin{align}
\mathcal{V}_{DM} = \frac{1}{4}\lambda_{41}\phi_1^4 + \frac{1}{4}\lambda_{42}\phi_2^4 +\frac{1}{4} \lambda_{412} \phi_1^2 \phi_2^2  + \frac{1}{4}\lambda_{31}\phi_1^3\phi_2 cos(3\theta_1 +\theta_2) +\frac{1}{4}\lambda_{32}\phi_1\phi_2^3 cos(\theta_1 - 3\theta_2).
\end{align}
After minimizing the potential for $3\theta_1 +\theta_2$ and $\theta_1 - 3\theta_2$ with $\phi_1 \neq 0$ and $\phi_2 \neq 0$, the above potential has the form, 
\begin{align}
\mathcal{V}_{DM} = \frac{1}{4}\lambda_{41}\phi_1^4 + \frac{1}{4}\lambda_{42}\phi_2^4 +\frac{1}{4} \lambda_{412} \phi_1^2 \phi_2^2  - \frac{1}{4}\lambda_{31}\phi_1^3\phi_2 -\frac{1}{4}\lambda_{32}\phi_1\phi_2^3.
\end{align}
Taking $X\equiv \frac{\phi_1}{\phi_2}$, the vacuum stability conditions become
\begin{align}
\lambda_{41} >0 ,    \lambda_{42} >0
\end{align}
and 
\begin{align}\label{leq0}
f(X_{min}) > 0
\end{align}
with
\begin{align}
f(X_{min}) =  \frac{1}{4}\lambda_{41}X^4 -\frac{1}{4}\lambda_{31}X^3 
+\frac{1}{4} \lambda_{412} X^2 - \frac{1}{4}\lambda_{32}X +\frac{1}{4}\lambda_{42},  
\end{align}
where $X_{min}$ labels the global minimum when $f^{'}(X_{min})=0$. Then, solving $f^{'}(X_{min})=0$ by the condition $f^{''}(X_{min}) > 0$, we obtain the third condition (\ref{leq0}) as

\begin{align}
\lambda_{41}X_{min}^3 - \frac{3}{4}\lambda_{31}X^2 +\frac{1}{2}\lambda_{412}X-\frac{1}{4}\lambda_{32} = 0,\label{CUB}
\\
\lambda_{42} > \frac{1}{4}\lambda_{31}X^3_{min}+\frac{1}{2}\lambda_{412}X^2_{min}+\frac{3}{4}\lambda_{32}X_{min}.
\end{align}
where
\begin{align}
X_{ min} = \Bigg\{
\begin{array}{ll}
\left(
P+\sqrt{P^2 + Q^3}    
\right)^{1/3}
+\left(
P-\sqrt{P^2 + Q^3}    
\right)^{1/3}\,, & D>0 \\
2\sqrt{-Q}\cos\left(
\frac{1}{3}\cos^{-1}\left(
\frac{P}{\sqrt{-Q^3}}
\right)
\right)
\,, & D<0
\end{array}
\end{align}
with $D \equiv P^2 + Q^3$, $P \equiv \frac{4\lambda_{31}\lambda_{41}\lambda_{412}-8\lambda_{32}\lambda_{41}^2-\lambda_{31}^3}{64\lambda_{41}^3}$, and $Q \equiv \frac{8\lambda_{41}\lambda_{412}-3\lambda^2_{31}}{48\lambda_{41}^2}$.

\subsection{Results}
\label{infAs}

In this section, we discuss how low energy constraints and the scalar power spectrum affect model parameters. As explained in the previous section, we consider singlet dark matter and mixed dark matter inflation respectively. We choose the following input parameters,
\begin{align}
    \xi_2/\xi_1,\xi_1,\lambda_{31},\lambda_{32},\lambda_{412},\lambda_{41},\lambda_{42}.
\end{align}
and fix $N$=62 in the analysis of scalar power spectrum.

\subsubsection{case A and B}
\begin{figure}[tbp]
\centering
\includegraphics[width=0.45\linewidth]{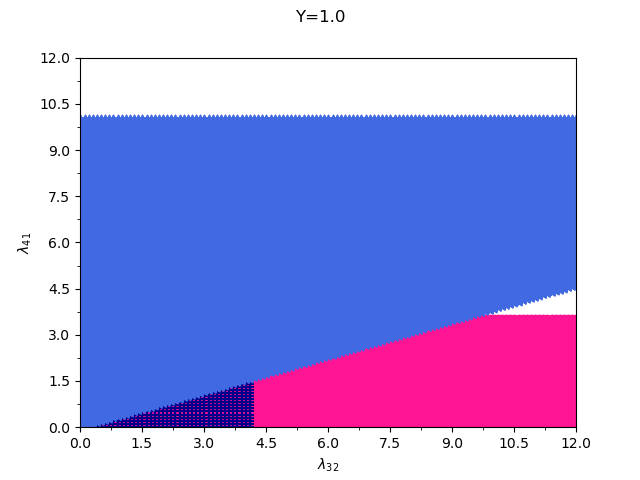}\,\,
\caption{Parameter space for singlet dark matter inflation with $\lambda_{412}=3\lambda_{41}$ at low-energy scale. Constraints from the model and scalar power spectrum. The blue region represents the vacuum stability, the dark blue and red region are favored due to perturbativity and unitarity respectively.}
\label{oneinf}
\end{figure}

In the case of $r \rightarrow 0$ or $r \rightarrow \infty$ scenario, singlet dark matter inflation takes effect under the condition $\xi_1\lambda_{412} >2\xi_2\lambda_{41}$ or $\xi_2\lambda_{412} > 2\xi_1\lambda_{42}$. According to the symmetry of $\xi_1 \leftrightarrow \xi_2$ and $\lambda_{41} \leftrightarrow \lambda_{42}$, Case B is equivalent to Case A, so only one of them needs to be studied and we select the latter. To simplify parameter space further, we choose $\lambda_{412} = 3\lambda_{41}$ and $\xi_1 \leq \xi_2$ so that $\xi_1\lambda_{412} >2\xi_2\lambda_{41}$ always holds. Then, there only three free parameters exist:{$\lambda_{32}, \lambda_{41}, \xi_1$}. We scan parameter space at $\lambda_{32}, \lambda_{41} \in[0.1,10]$ and $\xi_1 \in[10,10^5]$. Fig.(\ref{oneinf}) shows the allowed area restricted by vacuum stability, perturbativity, unitarity and scalar power spectrum. As we can see, perturbativity and scalar power spectrum put more limits on $\lambda_{32}$ and $\lambda_{41}$.
 
\subsubsection{case C}
\begin{figure}[tbp]
\centering
\includegraphics[width=0.32\linewidth]{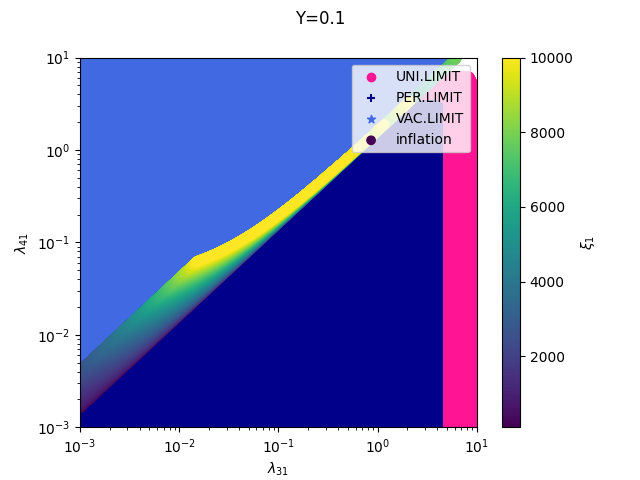}\,\,
\includegraphics[width=0.32\linewidth]{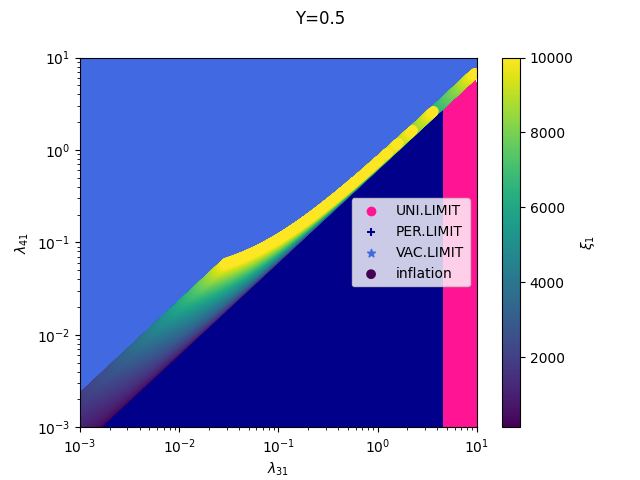}\,\,
\includegraphics[width=0.32\linewidth]{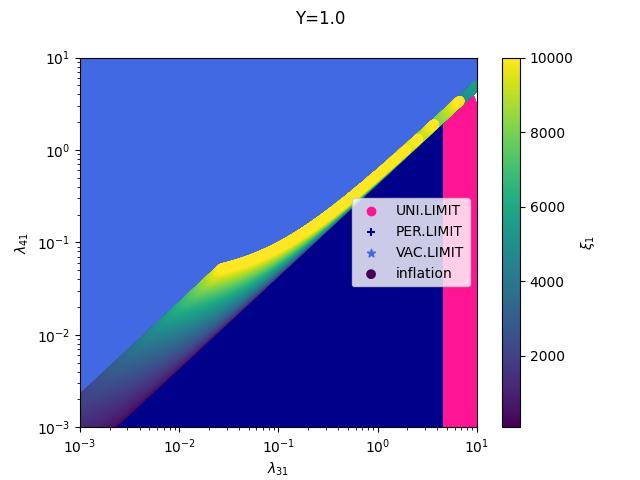}
\caption{parameter space for mixed dark matter inflation with $\lambda_{42}=Y^2\lambda_{41}$, $\lambda_{32}=Y\lambda_{31}$, $\lambda_{412}=2Y\lambda_{41}$ at $Y=\xi_2/\xi_1=1.0,0.5,0.1$. The parameters shown here are set at low-energy scale. The blue region represents the vacuum stability. The dark blue and red region are favored due to perturbativity and unitarity respectively.The histogram show the value of $\xi$.}
\label{mixv1}
\end{figure}

The mixed dark matter inflation takes place in the condition of $\frac{\partial {{\cal V}(r)}}{\partial r_c} =0$ and $\frac{\partial^2 {{\cal V}(r)}}{\partial r^2_c} > 0$ with a nonzero value of $\lambda_{31}$ and $\lambda_{32}$. The Eq.(\ref{casec1})-Eq.(\ref{casec3}) can be replaced by,
\begin{align}
 Y \equiv \xi_2/\xi_1, \lambda_{42}=Y^2\lambda_{41}, \lambda_{32}=Y\lambda_{31}, \lambda_{412}=2Y\lambda_{41}.   
\end{align}
We take ${\lambda_{31}, \lambda_{41},\xi_1}$ as free parameters and fix $Y$ with 0.1, 0.5, 1.0. The parameter space is sampled within $\lambda_{31}, \lambda_{41} \in[0.1,10]$ and $\xi_1 \in[10,10^5]$. Fig.(\ref{mixv1}) shows the viable range, bounded by vacuum stability, perturbativity, unitarity and scalar power spectrum, and the scalar power spectrum gives a stronger restriction on $\lambda_{31}$ and $\lambda_{41}$.


\section{Dark matter }
 \label{dm}
 
In this section, we discuss the dark matter phenomenology in the model. We focus on the case of mixed dark matter inflation and fix $Y=1.0, 0.5$ and $0.1$ respectively. In the $Z_5$ dark matter model, there are three $2\to 2$ processes related to dark matter, including annihilations, semi-annihilations and conversions. The Boltzmann equations related to the two-component dark matter are given by,
 \begin{eqnarray}
\label{boltzmann1}
\frac{dN_1}{dt} + 3H N_1&=&-\sigma_v^{1100}  \left(N_1^2-\bar{N}_1^2 \right) -
\sigma_v^{1120}\left( N_1^2- N_2 \frac{\bar{N}_1^2}{\bar{N}_2} \right)
- \sigma_v^{1122}\left( N_1^2- N_2^2 \frac{\bar{N}_1^2}{\bar{N}_2^2}
\right)\nonumber  \\
  && - \frac{1}{2}\sigma_v^{1112}\left( N_1^2- N_1 N_2 \frac{\bar{N}_1}{\bar{N}_2}\right)
      - \frac{1}{2}\sigma_v^{1222}\left( N_1 N_2- N_2^2\frac{\bar{N}_1}{\bar{N}_2}\right)\nonumber \\
               &&  -\frac{1}{2}\sigma_v^{1220}\left( N_1 N_2- N_2 \bar{N}_1 \right)
+\frac{1}{2}\sigma_v^{2210}(N_2^2-N_1\frac{\bar{N}_2^2}{\bar{N}_1})  \,, \\
\frac{dN_2}{dt}+ 3H N_2&=&-\sigma_v^{2200}  \left(N_2^2-\bar{N}_2^2 \right) -
\sigma_v^{2210}\left( N_2^2- N_1 \frac{\bar{N}_2^2}{\bar{N}_1} \right)
-\sigma_v^{2211}\left( N_2^2- N_1^2 \frac{\bar{N}_2^2}{\bar{N}_1^2}
\right)\nonumber  \\
  && - \frac{1}{2}\sigma_v^{2221}\left( N_2^2- N_1 N_2 \frac{\bar{N}_2}{\bar{N}_1}\right)
      - \frac{1}{2}\sigma_v^{1211}\left( N_1 N_2- N_1^2\frac{\bar{N}_2}{\bar{N}_1}\right)\nonumber \\
               &&  -\frac{1}{2}\sigma_v^{1210}\left( N_1 N_2- N_1 \bar{N}_2 \right)
+\frac{1}{2}\sigma_v^{1120}(N_1^2-N_2\frac{\bar{N}_1^2}{\bar{N}_2})   .      
\label{boltzmann2}
\end{eqnarray}
Where $N_{i}$ ($i=1,2$) are the number densities of $\phi_i$, and $\bar{N}_i$  the equilibrium values. $\sigma_v^{abcd}$ corresponds to the thermally averaged cross section of the $a+b\to c+d$ process, and the digits $0, 1, 2$ above the $\sigma_v$ represent different particles in the model, where $0$ is used for SM particles, $1$ represents $\phi_1$ or $\phi_1^\dagger$ and $2$ for $\phi_2$ or $\phi_2^\dagger$. To estimate the contribution of the three $2\to 2$ processes to the relic density of $\phi_1$, one can define the following three parameters for annihilation, semi-annihilation and conversion respectively:
 \begin{align}
    \chi_{an}&\equiv\frac{\sigma_v^{1100}}{\overline{\sigma_v^{1}}},\quad
    \chi_{se}\equiv\frac{\frac{1}{2}(\sigma_v^{1120}+\sigma_v^{1220}+\sigma_v^{1022})}{\overline{\sigma_v^{1}}},\quad
    \chi_{co}\equiv\frac{\sigma_v^{1122}+\sigma_v^{1112}+\sigma_v^{1222}}{\overline{\sigma_v^{1}}},\label{eq:semi}
\end{align}
with
\begin{align}
\overline{\sigma_v^{1}}\equiv\sigma_v^{1100}+\frac{1}{2}\sigma_v^{1120}+\sigma_v^{1122}+\sigma_v^{1112}+\sigma_v^{1222}+\frac{1}{2}\sigma_v^{1220}+\frac{1}{2}\sigma_v^{1022}.
\end{align}

We calculate the DM relic density with micrOMEGAs numerically \cite{Belanger:2014vza}, where we scan the parameter space of 
 \begin{eqnarray}
 M_1 \subset [1500\ \mathrm{GeV},1800\ \mathrm{GeV}],  M_2 \subset [1500 \rm \ GeV,3600\rm \ GeV]
 \end{eqnarray}
with $M_1$ the $\phi_1$ mass and $M_2$ the $\phi_2$ mass. We limit the parameter space with relic density constraint. On the other hand, we assume $M_1<M_2$ and $M_2<2M_1$ to guarantee both $\phi_1$ and $\phi_2$ the dark matter candidate. For different $Y$, we fix $\lambda_{S1}=\lambda_{S2}=0.75$ and  set other parameters constants respectively, the results are given as follows.
\begin{figure}[h]
\centering
\begin{minipage}[t]{0.48\textwidth}
\centering
\includegraphics[width=7cm,height=5cm]{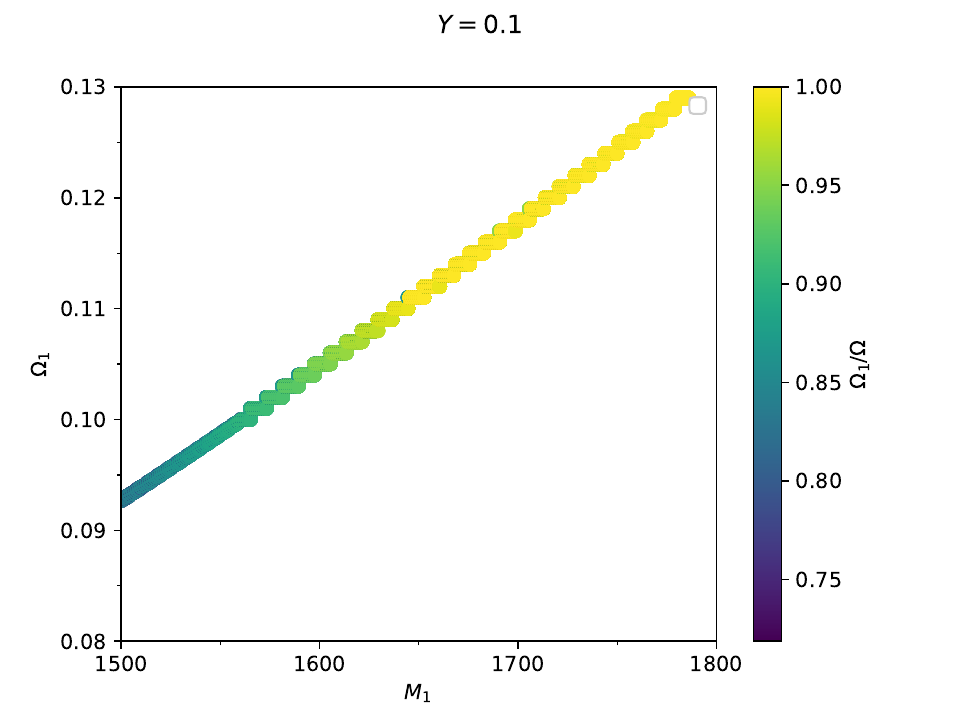}
\caption{Parameter space of $M_1-\Omega_1$ satisfying DM relic density constraint in the case of $Y=0.1$, and points with different colors  correspond to the fraction of $\phi_1$ to the total relic density.}
   \label{Fig:4}
\end{minipage}
\begin{minipage}[t]{0.48\textwidth}
\centering
\includegraphics[width=7cm,height=5cm]{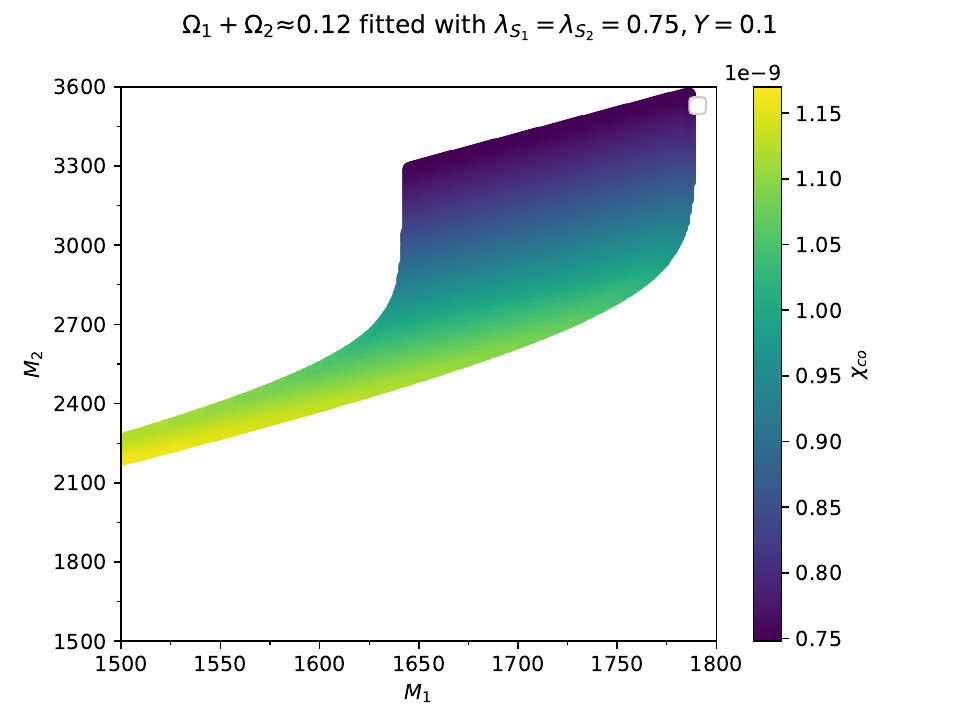}
\caption{Parameter space of $M_1-M_2$ satisfying DM relic density constraint in the case of $Y=0.1$, and points with different colors describe the fraction of conversion in the $2\to 2$ processes.}
  \label{Fig:5}
\end{minipage}
\end{figure}
\begin{figure}[h]
\centering
\begin{minipage}[t]{0.48\textwidth}
\centering
\includegraphics[width=7cm,height=5cm]{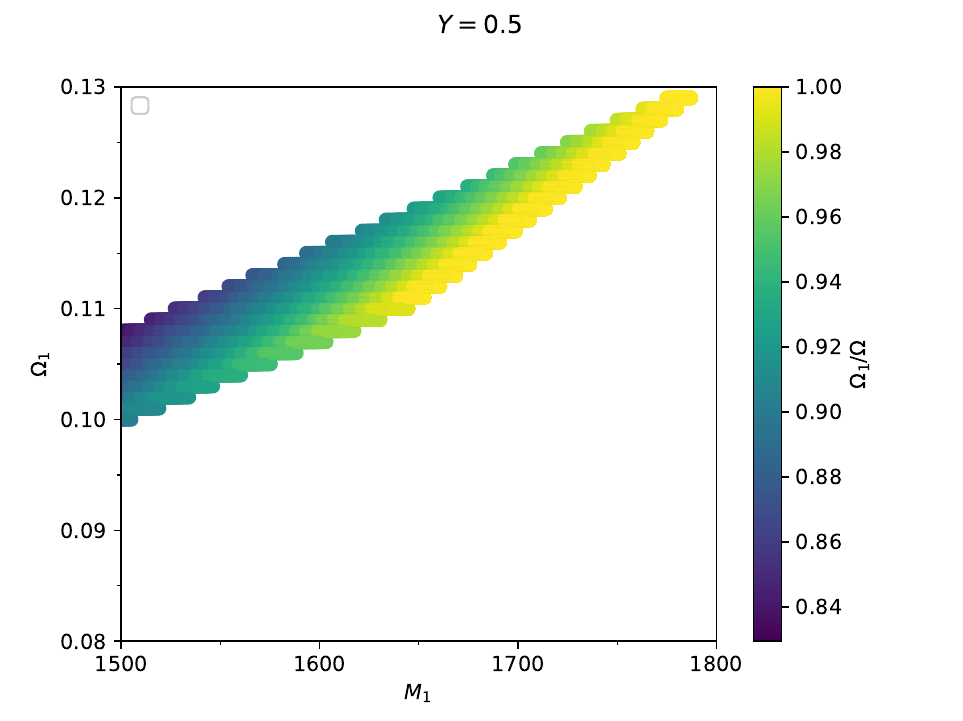}
\caption{Parameter space of $M_1-\Omega_1$ satisfying DM relic density constraint in the case of $Y=0.5$, and points with different colors  correspond to the fraction of $\phi_1$ to the total relic density.}
   \label{Fig:6}
\end{minipage}
\begin{minipage}[t]{0.48\textwidth}
\centering
\includegraphics[width=7cm,height=5cm]{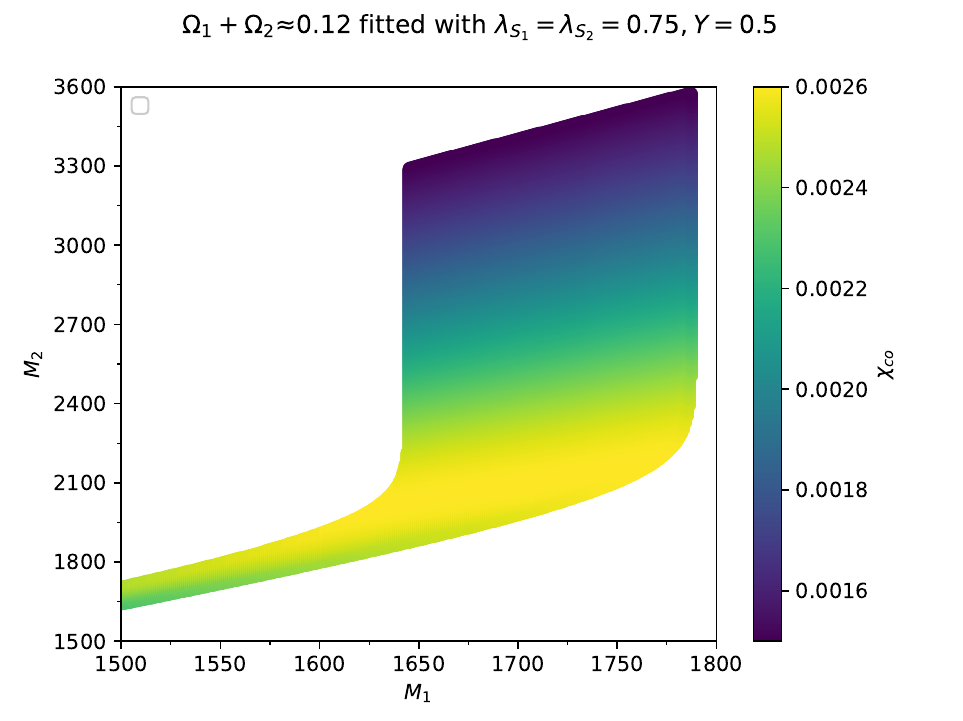}
\caption{Parameter space of $M_1-M_2$ satisfying DM relic density constraint in the case of $Y=0.5$, and points with different colors describe the fraction of conversion in the $2\to 2$ processes.}
  \label{Fig:7}
\end{minipage}
\end{figure}
\begin{figure}[h]
\centering
\begin{minipage}[t]{0.48\textwidth}
\centering
\includegraphics[width=7cm,height=5cm]{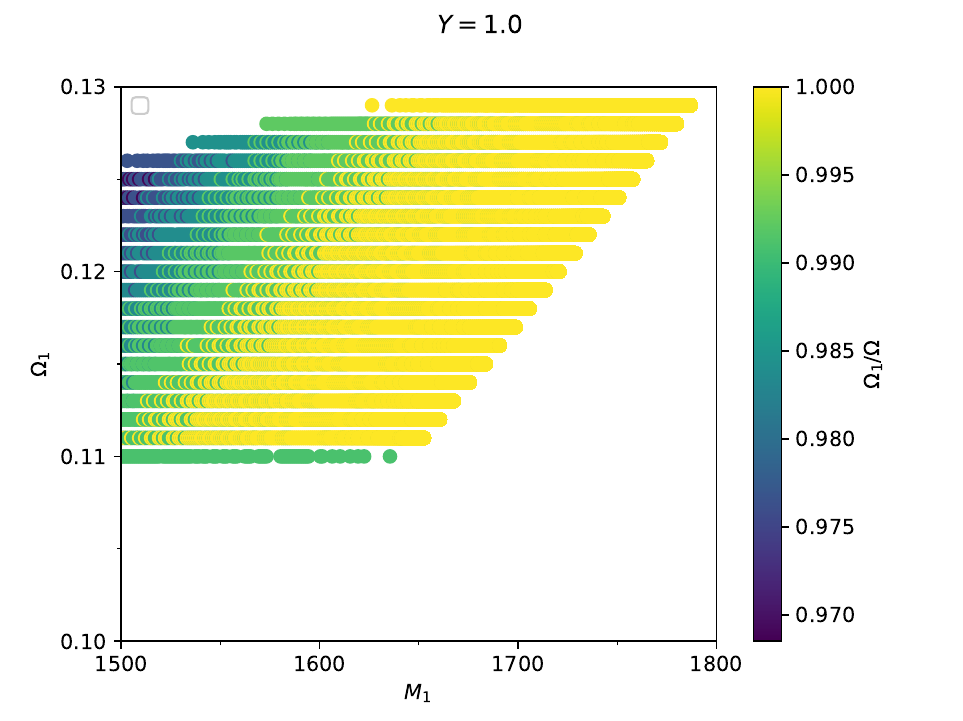}
\caption{Parameter space of $M_1-\Omega_1$ satisfying DM relic density constraint in the case of $Y=1$, and points with different colors  correspond to the fraction of $\phi_1$ to the total relic density.}
   \label{Fig:8}
\end{minipage}
\begin{minipage}[t]{0.48\textwidth}
\centering
\includegraphics[width=7cm,height=5cm]{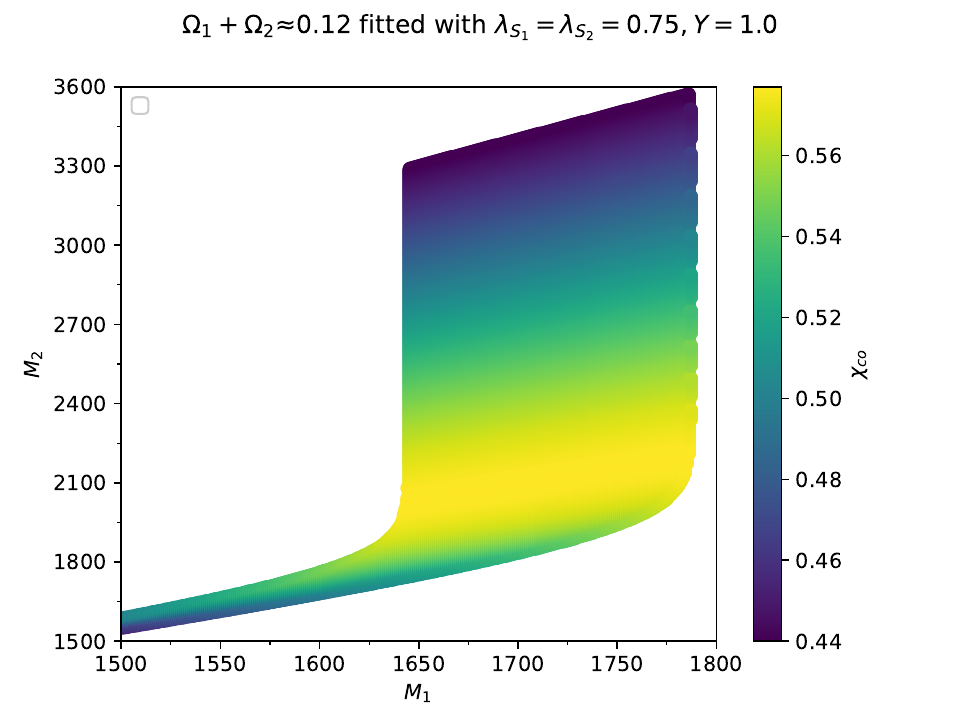}
\caption{Parameter space of $M_1-M_2$ satisfying DM relic density constraint in the case of $Y=1$, and points with different colors describe the fraction of conversion in the $2\to 2$ processes.}
  \label{Fig:9}
\end{minipage}
\end{figure}

In Fig.~\ref{Fig:4} and Fig.~\ref{Fig:5}, we fix $Y=0.1$, $\lambda_{41}=1$, $\lambda_{42}=Y^2\lambda_{41}$, $\lambda_{412}=0.002$, $\lambda_{31}=0.073030$ and $\lambda_{32}=0.000024$.
We give the relationship between $M_1$ and $\Omega_1$ satisfying relic density constraint in Fig.~\ref{Fig:4}, 
where points with different colors represent the fraction of $\phi_1$ in the total relic density and $\Omega_1$ is the relic density of $\phi_1$ while $\Omega$ is the total relic density.
The relic density of $\phi_1$ is stringently limited within a narrow region of $(0.09, 0.13)$, increasing with the increase of $M_1$.
In addition, the fraction of $\phi_1$ with $\Omega_1/\Omega$ is also increasing with the increase of $M_1$, and for $M_1 >1650$ GeV, the dark matter relic density is almost composed of $\phi_1$.
According to Fig.~\ref{Fig:5}, the viable region satisfying relic density constraint for $M_2$ is [2100, 3600]GeV, 
and we can have a more flexible parameter space for $M_2$ when $M_1>1650$ GeV since $\phi_2$ will be less relevant in the dark matter relic density.
In addition, Fig.~\ref{Fig:5} also gives the fraction of the conversion between $\phi_1$ and $\phi_2$ among the $2\to 2$ processes. 
The value of $\chi_{co}$ is at $10^{-9}$ level and can be negligible, which means the relic density is mainly determined by the annihilations as well as semi-annihilations. 

 In Fig.~\ref{Fig:6} and Fig.~\ref{Fig:7}, we discuss the relationship between $M_1-\Omega_1$ and $M_1-M_2$ in the case of $Y=0.5$, where we fix other parameters with  
$\lambda_{41}=1$, $\lambda_{42}= Y^2\lambda_{41}=0.25$, $\lambda_{412}=0.25$, $\lambda_{31}=0.816497$ and $\lambda_{32}=0.034021$. We have a more flexible value for $\Omega_1$ and $M_2$ compared with the case of $Y=0.1$, and the viable value for $M_2$ satisfying relic density constraint can decrease to 1600 GeV. Nevertheless, the contribution  of $\phi_1$ to the relic density is still dominant over $\phi_2$, and dark matter relic density is almost given by $\phi_1$ when $M_1$ is larger than 1700 GeV. 

  According to Fig.~\ref{Fig:8} and Fig.~\ref{Fig:9}, we give the results of $Y=1$, where we set $ \lambda_{42}=\lambda_{41}=1.01$, $\lambda_{412}=2.02$, $\lambda_{31}=2.332495$ and $\lambda_{32}=0.777498$. Similarly, $\phi_1$ is dominant on the dark matter relic density in Fig.~\ref{Fig:8}, although the conversion of $\phi_1$ to $\phi_2$ increases up to 0.56 as in Fig.~\ref{Fig:9}. Such a result  is consistent with the conclusion of ref.~\cite{Belanger:2020hyh} that the dark matter density is always dominated by the lighter component in the $Z_5$ model. 
 
  The elastic scattering of the dark matter off nuclei in the model arises from the Higgs portal interaction $\lambda_{Si}$ with $i=1, 2$, which can put a stringent constraint on the parameter space. The expression of the spin-independent (SI) cross section can be given by \cite{Belanger:2020hyh}. 
  \begin{eqnarray}
  \sigma_i= \frac{\lambda_{Si}^2}{4\pi}\frac{\mu_R^2m_p^2f_p^2}{m_H^4M_i^2}
    \label{ddeq}
  \end{eqnarray}
  where $i=1, 2$, $\mu_R$ is the reduced mass, $m_p$ is the proton mass, $m_H$ the SM Higgs mass and $f_p \approx 0.3$ is the quark content of the proton. Experiments on the direct detection of dark matter can be found in \cite{XENON:2018voc,PandaX-4T:2021bab}, and the PandaX-4T experiments \cite{PandaX-4T:2021bab} put the most stringent constraint on the spin-independent dark matter.

  

\section{Conclusion}
\label{sec:conclusions}

In this work, we consider the dark matter and inflation in the $Z_5$ model of the two-component dark matter. 
Two singlet scalars carrying different $Z_5$ charges are introduced to the SM as the dark matter candidate. 
Either one of the dark matter particles or both the two scalars can play the role of the inflaton, 
which corresponds to the case of singlet dark matter inflation and mixed dark matter inflation. 
 
We estimate the parameter space by considering all the theoretical constraints in the case of singlet dark matter inflation and mixed dark matter inflation respectively. 
We focus on the latter inflation case and then consider the dark matter relic density constraint. 
In the case of successful inflation, the viable parameter space not only satisfies correct relic abundance but is also limited by the direct detection result. 
It turns out that the dark matter density is dominated by the light component, which is consistent with the feature of the $Z_5$ model of the two-component dark matter.

\section*{Acknowledgements}
Part of the work is done by the author Aigeng Yang who died of illness and left us far away. 
Hao Sun is supported by the National Natural Science Foundation of China (Grant No. 12075043, No.12147205).

\appendix

\section{Unitarity constraints}
\label{uncon}

With the help of mutually unmixed sets of channels with definite charge, 
the full set of two-body scalar scalar scattering processes leads to $26 \times 26$ matrix which can be decomposed into four block submatrices. 
There are three submatrices with zero-charge channels in the initial/final states:
$S^{(1)}(10\times10)$, $S^{(2)}(5\times5)$, $S^{(3)}(9\times9)$.
The last one is $S^{(4)}(6\times6)$ with one-change channels. 

The first submatrix ${\cal M}_1$ corresponds to the scattering whose initial and final states are one of the following:
$(G^0\phi_1, G^0\phi_2, G^0\phi^{*}_1, G^0\phi^{*}_2, h\phi_1, h\phi_2, h\phi^{*}_1, h\phi^{*}_2, \phi_1\phi^{*}_2,\phi^{*}_1\phi_2)$.
From the vertex factors one can find
\begin{equation}
{\cal M}_1 =
\left(
\begin{array}{cccccccccc}
\frac{1}{2}\lambda_{S1}&0&0&0&0&0&0&0&0&0\\
0&\frac{1}{2}\lambda_{S2}&0&0&0&0&0&0&0&0\\
0&0&\frac{1}{2}\lambda_{S1}&0&0&0&0&0&0&0\\
0&0&0&\frac{1}{2}\lambda_{S2}&0&0&0&0&0&0\\
0&0&0&0&\frac{1}{2}\lambda_{S1}&0&0&0&0&0\\
0&0&0&0&0&\frac{1}{2}\lambda_{S2}&0&0&0&0\\
0&0&0&0&0&0&\frac{1}{2}\lambda_{S1}&0&0&0\\
0&0&0&0&0&0&0&\frac{1}{2}\lambda_{S2}&0&0\\
0&0&0&0&0&0&0&0&\lambda_{412}&0\\
0&0&0&0&0&0&0&0&0&\lambda_{412}
\end{array}
\right),
\end{equation}
We find that ${\cal M}_1$ has the following ten eigenvalues $e^i_j$:
\begin{align}
e^4_1=\frac{1}{2}\lambda_{S1},\\
e^4_2=\frac{1}{2}\lambda_{S2},\\
e^2_3=\lambda_{412}.
\end{align}

The second submatrix ${\cal M}_2$ corresponds to the scattering whose initial and final states are one of the following:
$(G^+G^-, \frac{G^0G^0}{\sqrt{2}}, \frac{hh}{\sqrt{2}}, \phi_1\phi^{*}_1, \phi_2\phi^{*}_2)$,
where the $\sqrt{2}$ accounts for identical particles. Again, from the vertex factors one can find
\begin{equation}
{\cal M}_2 =
\left(
\begin{array}{ccccc}
4\lambda_{H}&\sqrt{2}\lambda_{H}&\sqrt{2}\lambda_{H}&\lambda_{S1}&\lambda_{S2}\\
\sqrt{2}\lambda_{H}&3\lambda_{H}&\lambda_H&\frac{1}{\sqrt{2}}\lambda_{S1}&\frac{1}{\sqrt{2}}\lambda_{S2}\\
\sqrt{2}\lambda_H&\lambda_{H}&3\lambda_H&\frac{1}{\sqrt{2}}\lambda_{S1}&\frac{1}{\sqrt{S2}}\\
\lambda_{S1}&\frac{1}{\sqrt{2}}\lambda_{S1}&\frac{1}{\sqrt{2}}\lambda_{S1}&4\lambda_{41}&\lambda_{412}\\
\lambda_{S2}&\frac{1}{\sqrt{2}}\lambda_{S2}&\frac{1}{\sqrt{2}}\lambda_{S2}&\lambda_{412}&4\lambda_{42}
\end{array}
\right).
\end{equation}
We find that ${\cal M}_2$ has the following five eigenvalues:
\begin{align}
    e_i^1&=\frac{1}{2}X_i,(i=1,2,3)\\
    e_4^2 &= 2\lambda_{H}.
\end{align}
Where $X_i$ denotes the roots of the cubic equation as follows:
\begin{align}\nonumber
 &X^3 - 4(2\lambda_{41}+2\lambda_{42}+3\lambda_{H})X^2 \\
 &+ 4(16\lambda_{41}\lambda_{42}-\lambda^2_{412}+24\lambda_{41}\lambda_{H}+24\lambda_{42}\lambda_{H}-2\lambda^2_{S1}-2\lambda^2_{S2})X  \nonumber \\
 &-16(48\lambda_{41}\lambda_{42}\lambda_{H}-3\lambda^2_{412}\lambda_{H}-4\lambda_{42}\lambda^2_{S1}+2\lambda_{412}\lambda_{S1}\lambda_{S2}-4\lambda_{41}\lambda^2_{S2})=0.
\end{align}
We apply the Samuelson's inequality~\cite{Samine} to find the region of real roots $X_i$,
i.e. for the polynomial 
\begin{eqnarray}
a_nx^n+a_{n-1}x^{n-1}+a_1x+a_0
\end{eqnarray}
with only real roots, the roots lie in the interveral bounded by
\begin{align}
x_{\pm}=-\frac{a_{n-1}}{na_n}\pm\frac{n-1}{na_n}\sqrt{a_{n-1}^2-\frac{2n}{n-1}{a_n}{a_{n-2}}}.    
\end{align}

The third submatrix ${\cal M}_3$ corresponds to the scattering whose initial and dinal states are one of the following:
$(hG^0, \frac{1}{\sqrt{2}}\phi_1\phi_1, \frac{1}{\sqrt{2}}\phi^*_1\phi^*_1, \frac{1}{\sqrt{2}}\phi_2\phi_2, \frac{1}{\sqrt{2}}\phi^*_2\phi^*_2, \phi_1\phi_2, \phi^*_1\phi^*_2, \phi_1\phi^*_2, \phi^*_1\phi_2)$.
One can find
\begin{equation}
{\cal M}_3 =
\left(
\begin{array}{ccccccccc}
     2\lambda_H&0&0&0&0&0&0&0&0\\
     0&2\lambda_{41}&0&0&0&0&\frac{3}{\sqrt{2}}\lambda_{31}&0&0\\
     0&0&2\lambda_{41}&0&0&\frac{3}{\sqrt{2}}\lambda_{31}&0&0&0\\
     0&0&0&2\lambda_{42}&0&0&0&\frac{3}{\sqrt{2}}\lambda_{32}&0\\
     0&0&0&0&2\lambda_{42}&0&0&0&\frac{3}{\sqrt{2}}\lambda_{32}\\
     0&0&\frac{3}{\sqrt{2}}\lambda_{31}&0&0&\lambda_{412}&0&0&0\\
     0&\frac{3}{\sqrt{2}}\lambda_{31}&0&0&0&0&\lambda_{412}&0&0\\
     0&0&0&\frac{3}{\sqrt{2}}\lambda_{32}&0&0&0&\lambda_{412}&0\\
     0&0&0&0&\frac{3}{\sqrt{2}}\lambda_{32}&0&0&0&\lambda_{412}
\end{array}
\right).
\end{equation}
${\cal M}_3$ has the following nine eigenvalues:
\begin{align}
e_{1,2}^4 &=\frac{1}{2}\left(2\lambda_{41}+\lambda_{412}\pm\sqrt{18\lambda_{31}^2+4\lambda^2_{41}-4\lambda_{41}\lambda_{412}+\lambda^2_{412}}\right),\\
e_3 &=2\lambda_H.
\end{align}
The last submatrix $\mathcal{M}_4$ corresponds to the one-change channels occur for two-by-two body scattering 
between the six charged states $(hG^+, G^0G^+, \phi_1G^+, \phi_2G^+, \phi^*_1G^+, \phi^*_2G^+)$.
We have ${\cal M}_4$ and its eigenvalues as follows:
\begin{equation}
    {\cal M}_4 =
    \left(
    \begin{array}{cccccc}
        2\lambda_H&0&0&0&0&0\\
         0&2\lambda_H&0&0&0&0\\
         0&0&\lambda_{S1}&0&0&0\\
         0&0&0&\lambda_{S2}&0&0\\
         0&0&0&0&\lambda_{S1}&0\\
         0&0&0&0&0&\lambda_{S2}
    \end{array}
    \right),
\end{equation}
\begin{align}
    e^2_1&=2\lambda_H,\\
    e^2_2&=\lambda_{S1},\\
    e^2_3&=\lambda_{S2}.
\end{align}

\section{Renormalization Group Equations}
To properly connect inflation scale with low energy scale of the dark matter physics, one should consider the RGEs of the relevant parameters. 
We obtain the Renormalization Group Equations (RGEs) of the model with SARAH~\cite{Staub:2013tta} at one-loop level, 
and the beta functions of the relavant quartic couplings are given by,
{\allowdisplaybreaks  
\begin{align} 
\beta_{\lambda_{31}} & =  
6 \lambda_{31} \Big(2 \lambda_{41}  + \lambda_{412}\Big)\\
\beta_{\lambda_{32}} & = 6 \lambda_{32} \Big(2 \lambda_{42}  + \lambda_{412}\Big)\\ 
\beta_{\lambda_{H}} & =  
+\frac{27}{200} g_{1}^{4} +\frac{9}{20} g_{1}^{2} g_{2}^{2} +\frac{9}{8} g_{2}^{4} -\frac{9}{5} g_{1}^{2} \lambda_{H} -9 g_{2}^{2} \lambda_{H} +24 \lambda_{H}^{2} +\lambda_{S1}^{2}+\lambda_{S2}^{2} \nonumber \\ 
 & +12 \lambda_{H}y_t^2 -6y_t^4\\
\beta_{\lambda_{42}}& =  
20 \lambda_{42}^{2}  + 2 \lambda_{S2}^{2}  + \frac{9}{2} |\lambda_{32}|^2  + \lambda_{412}^{2}\\ 
\beta_{\lambda_{412}}& =  
4 \Big(2 \lambda_{412} \lambda_{42}  + 2 \lambda_{41} \lambda_{412}  + \lambda_{S1} \lambda_{S2}  + \lambda_{412}^{2}\Big) + 9 |\lambda_{31}|^2  + 9 |\lambda_{32}|^2 \\ 
\beta_{\lambda_{41}} & =  
20 \lambda_{41}^{2}  + 2 \lambda_{S1}^{2}  + \frac{9}{2} |\lambda_{31}|^2  + \lambda_{412}^{2}\\
\beta_{\lambda_{H}}& =  
+\frac{27}{200} g_{1}^{4} +\frac{9}{20} g_{1}^{2} g_{2}^{2} +\frac{9}{8} g_{2}^{4} -\frac{9}{5} g_{1}^{2} \lambda_{H} -9 g_{2}^{2} \lambda_{H} +24 \lambda_{H}^{2} +\lambda_{S1}^{2}+\lambda_{S2}^{2} \\ 
 & +12 \lambda_{H}y_t^2  -6y^4_t  \\
 \beta_{\lambda_{S2}} & =  
+2 \lambda_{412} \lambda_{S1} -\frac{9}{10} g_{1}^{2} \lambda_{S2} -\frac{9}{2} g_{2}^{2} \lambda_{S2} +8 \lambda_{42} \lambda_{S2} +12 \lambda_{H} \lambda_{S2} +4 \lambda_{S2}^{2} 
 +6 \lambda_{S2}y_t^2 \\ 
 \beta_{\lambda_{S1}} & =  
-\frac{9}{10} g_{1}^{2} \lambda_{S1} -\frac{9}{2} g_{2}^{2} \lambda_{S1} +8 \lambda_{41} \lambda_{S1} +12 \lambda_{H} \lambda_{S1} +4 \lambda_{S1}^{2} +2 \lambda_{412} \lambda_{S2}  
 +6 \lambda_{S1}y_t^2
\end{align}
} 
where $y_t$ is the Top Yukawa coupling. For the trilinear scalar couplings, the RGEs are given by,
{\allowdisplaybreaks  
\begin{align} 
\beta_{\mu_{S2}} & =  
4 \Big(\lambda_{412} + \lambda_{42}\Big)\mu_{S2}  + 6 \lambda_{31} \mu_{S1}^*  + 6 \mu_{S1} \lambda_{32}^* \\ 
\beta_{\mu_{S1}} & =  
4 \lambda_{412} \mu_{S1}  + 4 \lambda_{41} \mu_{S1}  + 6 \lambda_{31} \mu_{S2}^*  + 6 \lambda_{32} \mu_{S2} .
\end{align}
}

\bibliographystyle{JHEP}
\bibliography{references.bib}
\end{document}